\documentclass[aps,prl,twocolumn,groupedaddress,amsmath,amssymb,showpacs]{revtex4}
\usepackage{amssymb}
\usepackage{amsmath}
\usepackage{graphicx}
\usepackage{epsfig}
\usepackage{epstopdf}
\usepackage{dcolumn}
\usepackage{bm}

\begin{document}

\preprint{Contact v.4}
\title{Theory of Spin-Dependent Phonon-Assisted Optical Transitions in Silicon}
\author{Pengke Li} \email{pengke@ece.rochester.edu}
\author{Hanan Dery}
%\date{\today}
\affiliation{Department of Electrical and Computer Engineering,
University of Rochester, Rochester, New York, 14627}
\begin{abstract}
A theory for the relation between the spin polarization and luminescence in silicon is presented. The theory provides intuitive relations for phonon-assisted optical transitions between the conduction and valence band edges. It is shown that an opposite behavior of longitudinal and transverse optical phonon-assisted transitions is responsible to recent experimental results of spin injection into silicon. The effects of spin-orbit coupling and doping are studied in detail.
\end{abstract}
%\keywords{spintronics, polarized electroluminescence, Spin-orbit coupling in silicon}
\maketitle

Optical orientation and luminescence polarization are widespread techniques in studying the spin of electrons in direct gap semiconductors \cite{OO_84,Zutic_RMP04}. These techniques, however, are not straightforward in indirect gap semiconductors where optical transitions are mediated by electron-phonon interactions \cite{Lampel_PRL68,Asnin_JETP76,Pikus_SPSS77,Nastos_PRB07}. In spite of a recent important progress in  spin injection into silicon \cite{Appelbaum_Nature07,Jonker_NaturePhysics07,Kioseoglou_APL09,Li_APL09,Grenet_APL09,Zutic_PRL06,Huang_APL08,Dash_Nature09,Jansen_NatureMater10}, there is no theory that enables a quantitative analysis of the luminescence from spin injected silicon. Such knowledge can be used to accurately determine the electrons' spin polarization from the measured circular polarization \cite{Jonker_NaturePhysics07,Kioseoglou_APL09,Li_APL09,Grenet_APL09}. Similarly, it can be used to infer the spin injection efficiency across ferromagnet/silicon interfaces \cite{Appelbaum_Nature07,Jonker_NaturePhysics07,Kioseoglou_APL09,Li_APL09,Grenet_APL09,Zutic_PRL06,Huang_APL08,Dash_Nature09,Jansen_NatureMater10}.

In this letter we study the relation between the spin polarization of electrons in silicon and the circular polarization degree of the luminescence. By invoking symmetry arguments, we first provide concise  selection rules for each of the phonon-assisted optical transitions in an unstrained bulk silicon. Then in the main part, these rules help in analyzing the polarized luminescence spectrum calculated by a comprehensive rigid-ion model. Our theoretical results elucidate recent experiments of spin injection from iron to silicon \cite{Jonker_NaturePhysics07,Kioseoglou_APL09,Li_APL09}. 
\begin{figure} [h]
\includegraphics[width=8.5 cm]{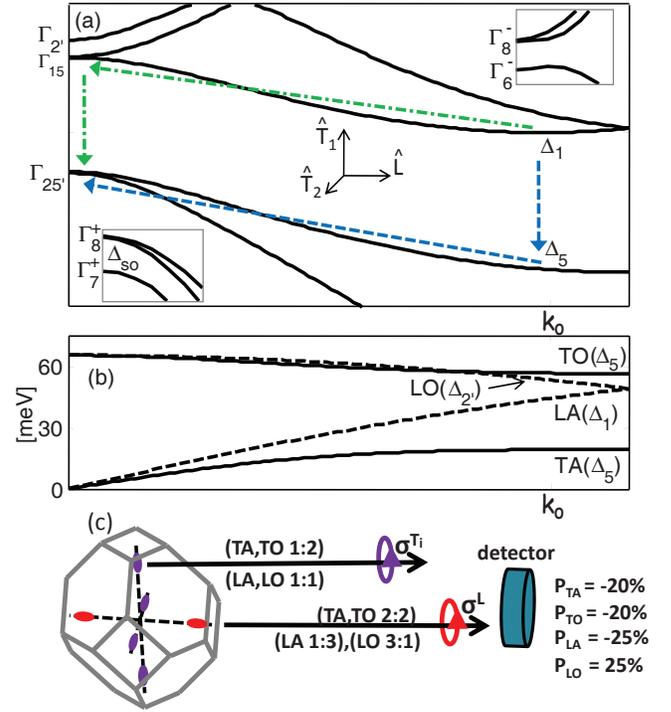}
\caption{(a) Band structure and symmetry notations of silicon along the $\Delta$ axis. The dot-dashed and dashed arrows represent the two dominant contributions for phonon-assisted optical transitions. The lower left (upper right) inset magnifies the splitting of the bands due to spin-orbit coupling  close to the $\Gamma_{25'}$ ($\Gamma_{15}$) zone center region.  (b) Phonon branches along the $\Delta$ axis and their corresponding symmetries. (c) Phonon dependent circular polarization properties for optical transitions of spin-up electrons from valleys whose axis lies along ($\sigma^L$) or perpendicular ($\sigma^T$) to the light propagation.
} \label{Fig:band}
\end{figure}

The luminescence of free carriers in bulk silicon involves transitions between electrons from the six equivalent conduction band valleys along the $\Delta$-symmetry axes and heavy or light holes at the top of the valence band (Fig.~\ref{Fig:band}(a)). Due to the crystal translational symmetry and the indirect absorption edge, phonon emission or absorption are required to offset a crystal momentum difference of $k_0$$\,$$\simeq$$\,$$0.85$$\,$$\times$$\,$$(2\pi/a)$ where $a$=5.43$\AA$ is the lattice constant. The light intensity is proportional to,
\begin{eqnarray}
I_{\hat{\mathbf{e}},\ell} \!\propto \! \left| \sum_n \! \frac{\langle f|H_R^{\hat{\mathbf{e}}}|n\rangle \langle n|H_{e-i}^\ell|i\rangle}{E_i-E_n-\hbar\omega_\ell}\!+\! \frac{\langle f|H_{e-i}^\ell|n\rangle\langle n| H_R^{\hat{\mathbf{e}}} |i\rangle }{E_i-E_n-\hbar\omega_{0}}\right|^2\!\!.
\label{eq:second order}
\end{eqnarray} %I_{\sigma_{\pm}}^l \propto
$H_R^{\hat{\mathbf{e}}}$ and $H_{e-i}^\ell$ related matrix elements denote, respectively, radiation-matter and electron-phonon interactions where $\hat{\mathbf{e}}$ is the light polarization vector and $\ell$ is the phonon mode: transverse-acoustic (TA), transverse-optical (TO), longitudinal-acoustic (LA) or longitudinal-optical (LO). The angular frequency of the photon is $\omega_0$ and of the phonon is $\omega_\ell$. The initial state at the conduction band is given by $|i\rangle$ and the final state at the valence band by $|f\rangle$. The first and second terms involve, respectively, optical transition paths with intermediate states, $|n\rangle$, whose crystal momentum is zero or $k_0$. The dot-dashed and dashed arrows in Fig.~\ref{Fig:band}(a) show the important respective paths
where the vertical (non-vertical) arrows denote the photon (phonon) interaction. The circular polarization of the spin-dependent luminescence strongly depends on the phonon dispersion and symmetries along the $\Delta$-axis. This information is shown in Fig.~\ref{Fig:band}(b). The TA and TO phonons share the same symmetry and therefore to first order their associated optical transitions are destined to  have similar polarization. However, it will be shown that this symmetrical behavior breaks down due to the proximity of the LO and TO phonon branches around $k_0$.

\begin{figure}
\includegraphics[width=8cm]{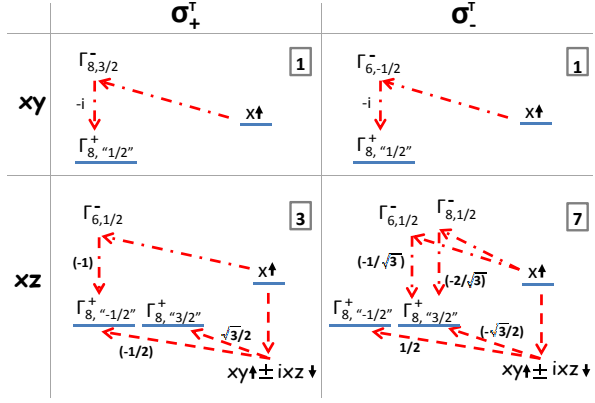}
\caption{ Transverse-phonon-assisted optical transitions of spin-up electrons from the $\hat{\mathbf{x}}$ valley. The upper and lower diagrams correspond to different phonon modes whereas the left and right diagrams to circular polarization of $\hat{\mathbf{x}} \pm i \hat{\mathbf{y}}$. The numbers next to the arrows denote the relative amplitude of the corresponding transition. Note that if there are several paths to reach the same final state then the interference is constructive (numbers in brackets have the same sign). The relative contribution of each phonon and photon configuration is indicated at the top right corner of each diagram. The circular polarization ratio ($\sigma_+^T:\sigma_-^T$) is 1:2.
} \label{Fig:TA}
\end{figure}

Using the symmetries of the electronic states and of the phonon modes, one can derive concise polarization ratios between the left and right circularly polarized luminescence. Figure~\ref{Fig:band}(c) lists the phonon-assisted circular polarization ratios and the ensuing total circular polarizations. These ratios correspond to transitions of spin-up electrons and they depend on whether electrons reside in valleys whose $\mathbf{k}_0$ is parallel or perpendicular to the light propagation axis (which is also the spin quantization axis). Figure~\ref{Fig:TA} shows the corresponding transverse-phonon-assisted optical transition diagrams from the $\hat{\mathbf{x}}$ valley (perpendicular to light propagation). Details about the construction of this and similar diagrams are found in the auxiliary material \cite{auxiliary}.

The interference effects between the various intermediate states is crucial for the analysis of the circular polarization. We focus on the resulting opposite polarization of the LO and the other phonon-assisted optical modes (Fig.~\ref{Fig:band}(c)). This suggests contributions of opposite sign to the circular polarization of emitted photons whose energy is about $E_g\!-\!\hbar\omega_{op}$  where $E_g$ is the gap energy and $\omega_{op}\approx \omega_{LO} \approx \omega_{TO}$ (due to the nearly degenerate energies of the LO and TO phonons). These photons will have smaller circular polarization compared with photons whose energy is around $E_g\!-\!\hbar\omega_{TA}$ in spite of the similar symmetry of TA and TO phonons. Other effects of the interference are also instructive for analyzing the luminscence spectrum. First, the interference of zone center and $k_0$ intermediate states is constructive (destructive) for transverse (longitudinal) phonons \cite{Glembocki_PRL82,Bednarek_PRL82}. This is the reason for the negligible intensity of the LA phonon-assisted optical transition. The symmetry of LO phonons, however, precludes transitions to the $\Gamma_{15}$ intermediate states \cite{Lax_PR61}. Consequently, the LO phonon-assisted optical transition and its circular polarization are dominated by intermediate states below the gap. The destructive interference with symmetry allowed intermediate states above the gap (e.g., the $\Gamma_{2'}$) is not sufficient to destroy the LO phonon-assisted transition or to significantly alter its associated circular polarization.

The robustness of the circular polarization and the relative intensity of the various spectral lines are calculated using realistic modeling of Eq.~(\ref{eq:second order}). The electronic states and energies are calculated via an empirical local pseudopotential method with spin-orbit coupling \cite{Cheli_PRB76}. The latter generates the correct energy splitting between $\Gamma_{8}^{+}$ and $\Gamma_{7}^{+}$ ($\Delta_{so}$=44~meV). The phonon modes and their dispersion are obtained by the adiabatic bond charge model \cite{Weber_PRB77}. The calculated energy band structure and phonon dispersion curve along the $\Delta$-axis are shown, respectively, in Figs.~\ref{Fig:band}(a)~and~(b). The radiation-matter matrix elements are calculated using the electric dipole approximation. The electron-phonon matrix elements are calculated via the rigid-ion approximation \cite{Allen_PRB81},
\begin{eqnarray}
\langle \varphi_{\mathbf{k}_f,m_f}|H_{e-i}^{\ell}| \varphi_{\mathbf{k}_i,m_i} \rangle \nonumber =\!\!\!  &A& \!\!\!\sum_{\mathbf{g}_1,\mathbf{g}_2} \! \sum_{\chi_1,\chi_2} \!\! V_{\mathbf{q}} C_{\mathbf{g}_2,\chi_2}^{\mathbf{k}_i,m_i} \! \left(C_{\mathbf{g}_1,\chi_1}^{\mathbf{k}_f,m_f}\right)^\ast   \nonumber \\ \scriptstyle{\times} \displaystyle \lbrace \mathbf{q} \cdot \mathbf{u}_{+}^{\ell}\cos(\Delta \mathbf{g}\cdot \boldsymbol{\tau}) &+& \mathbf{q} \cdot \mathbf{u}_{-}^{\ell}\sin(\Delta \mathbf{g}\cdot \boldsymbol{\tau}) \rbrace
\label{eq:pseudopotential}
\end{eqnarray}
The wavefunctions are taken from the pseudopotential model and given by $\varphi_{\mathbf{k},m}=\sum_{\mathbf{g}_j,\chi_j} C_{\mathbf{g}_j,\chi_j}^{\mathbf{k},m} \exp(i(\mathbf{g}_j+\mathbf{k})\cdot \mathbf{r})$ where  $\mathbf{k}$, $\mathbf{g}$, $m$ and $\chi$ denote, respectively, the wavevector, reciprocal lattice wavevector, band index and spin state. $V_{\mathbf{q}}$ is the pseudopotential of wavevector $\mathbf{q}=\Delta \mathbf{k}-\Delta\mathbf{g}$ where $\Delta \mathbf{k} = \mathbf{k}_f-\mathbf{k}_i$ and $\Delta\mathbf{g} = \mathbf{g}_1-\mathbf{g}_2$.  $A$=$i\sqrt{\hbar/(2M\omega_{\ell,\scriptstyle{\Delta \mathbf{k}}}\displaystyle)}$ where $M$ is the mass of a silicon atom. The phonon displacement vectors, $\mathbf{u}_{+}^{\ell}$ and $\mathbf{u}_{-}^{\ell}$, are the ``in-phase'' and ``out-of-phase'' motion of the two silicon atoms in the unit cell, respectively. These vectors are calculated via the adiabatic bond charge model. The equilibrium atom positions relative to the origin (midpoint) are given by $\pm \boldsymbol{\tau}$ where $\boldsymbol{\tau}=(a,a,a)/8$. The effect of the spin-orbit potential during a virtual transition is negligible and thus we consider only the local pseudopotential part in phonon-assisted transitions to intermediate states. We have verified that the Elliot-Yafet spin-flip mechanisms and their interference are of minor importance in our case of interest \cite{Yafet_1963, Cheng_PRL10}.

\begin{table} \caption{
Relative light intensities ($I_r$$=$$I_{\sigma_+}$$+$$I_{\sigma_-}$) and circular polarization degrees ($P$$=$$($$I_{\sigma_+}$$-$$I_{\sigma_-}$$)$$/$$I_r$) due to transitions with spin-up electrons. The light propagates along the $+z$ direction. The light intensities are normalized with respect to the TO intensity from the $x$ or $y$ valleys. The total intensities are the sum of contributions from all six valleys. The LA phonon results are not shown (negligible intensity).
}\label{tab:table1}
%%\begin{ruledtabular}
\renewcommand{\arraystretch}{1.5}
\begin{tabular}{l|ccc|ccc|ccc}
\hline \hline
valley  & &   $x$,$y$   &       &    &   $z$    &    &    & total &     \\ \hline \hline
mode  &TO &   LO  & TA    & TO &    LO   & TA   & TO &   LO  & TA  \\ \hline
$I_r$ & 1 & 0.115 & 0.086 & 1.41 & 0.23 & 0.092& 6.82 & 0.92&0.53\\ \hline
$P$[\%]&-32.3& 5.3 & -36   & 0.01 &   50.1 & 0.7   & -18.8& 27.7 &-23.5 \\ \hline \hline
\end{tabular}
%\end{ruledtabular}
\end{table}

Table~\ref{tab:table1} lists numerical results of relative light intensities and their circular polarization due to transitions of spin-up electrons at the bottom of the conduction band to heavy and light hole states at the top of the valence band. The left (middle) column corresponds to the contribution from each of the four (two) conduction band valleys whose direction is perpendicular (parallel) to the light propagation direction. The relative total intensities from all six valleys (right column) are consistent with the luminescence spectrum in silicon \cite{Wagner_PRB84}. The polarization values are consistent with those derived by symmetry arguments (Fig.~\ref{Fig:band}(c)). The non-zero polarization of the LO mode from perpendicular valleys ($\sim$$5\%$) is due to the small effect of transition paths via the $\Gamma_{2'}$ conduction band. In addition, the small deviation of the TA and TO total polarizations from 20\% (as predicted by symmetry arguments) generally comes from differences in the transition intensities of zone center and $k_0$ intermediate paths. The pseudopotential interpolation technique ($V_{\mathbf{q}}$ in Eq.~(\ref{eq:pseudopotential})), and the origin for the different intensities of TO and TA related optical transitions are spin independent and given in the auxiliary information \cite{auxiliary}.

\begin{figure}
\includegraphics[height=7cm,width=9cm]{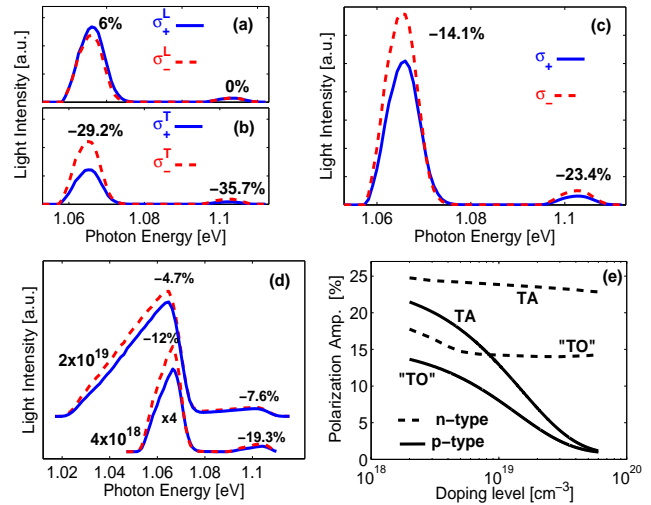}
\caption{
Calculated polarized spectra in doped silicon at 10~K. (a)-(c) show results of 4$\times$10$^{18}$~cm$^3$ n-type silicon. (a)/(b) show the contribution from valleys whose axis is along/perpendicular to the light propagation direction. (c) shows the contribution from all six valleys. (d) shows the polarized spectra for 4$\times$10$^{18}$~cm$^3$ and 2$\times$10$^{19}$~cm$^3$ p-type silicon. The former is magnified for clarity. The spectra in (a)-(d) consider the energy gap reduction with doping \cite{Wagner_PRB84}. (e) shows the polarization amplitude at the TA peak and TO dominated peak versus doping level (at the metallic regime). } \label{Fig:spectrum}
\end{figure}

The spectral width of the circularly polarized region is strongly affected by doping. The previous results were derived for transitions between the extremal points of the conduction and valence bands. However, in a p-type (n-type) doped silicon one should consider all of the states in the valence (conduction) band that can take part in the transition as well as the ensuing larger set of intermediate states. We have integrated the transition probability of the various states in the Brillouin zone weighted by Fermi distribution at $T=$10~$K$. Fig.~\ref{Fig:spectrum}(a)-(c) shows the resulting spectra of $\sigma_+$ and $\sigma_-$ for light propagation along the $+z$ direction in 4$\times$10$^{18}$~cm$^3$ n-type silicon. (a)-(c) denote, respectively, the contribution from the $z$, $x$ and all six valleys. The dominant peak just above 1065~meV originates predominately from the TO phonon-assisted transitions. Their associated circular polarization signal comes from the four transverse valleys ($\pm x$,~$\pm y$). About 15\% of the dominant peak originates from LO associated transitions which are responsible for the small polarization of the longitudinal valleys ($\pm z$). The opposite sign contributions to the total circular polarization (Fig.~\ref{Fig:spectrum}(c)) results in an evident smaller polarization degree of the dominant peak compared with the less intense TA associated peak (around 1105~meV). Figure~\ref{Fig:spectrum}(d) shows the total spectra in cases of 4$\times$10$^{18}$~cm$^3$ and 2$\times$10$^{19}$~cm$^3$ p-type silicon. We notice two features when comparing the n-type and p-type spectra at a doping level of 4$\times$10$^{18}$~cm$^3$ (Figs.~\ref{Fig:spectrum}(c) \& (d)). First, the p-type peaks are broader due to the position of the Fermi level; $E_F$$\sim$16~meV below the valence band edge vs $E_F$$\sim$8~meV above the multi-valley conduction band edge. Second, the p-type spectrum is less polarized due to stronger mixing of hole states away from the $\Gamma$-point. Both the broadening and reduced polarization are evident in the spectrum of the 2$\times$10$^{19}$~cm$^3$ p-type silicon. Finally, we can identify a common rule shared by the spectra of all doping cases in the metallic regime ($\gtrsim$2$\times$10$^{18}$~cm$^3$). The polarization has its largest amplitude at the lower energy edge of a given phonon-assisted transition. At this photon energy, the initial and final states are from the extrema states whereas at higher energies the mixing of states plays a key role in lowering the polarization.

The effect of spin-orbit coupling is most evident at heavily p-type silicon where the Fermi level is close to (or below) the split-off band. The proximity of the split-off band renders the holes state mixing effective already $\sim$10~meV away from the $\Gamma$-point. Fig.~\ref{Fig:spectrum}(e) shows the polarization values at the TA peak and TO dominated peak versus doping levels in the metallic regime. In n-type silicon, the polarization reduces only slightly with doping concentration. The initial polarization reduction of the TO dominated peak is due to the merging with the LO peak. This is essentially a broadening effect and it is effective when $E_F$$>$$\hbar\omega_{TO}$-$\hbar\omega_{LO}$$\approx$4~meV above the conduction band edge ($>$4$\times$10$^{18}$~cm$^3$). In p-type cases, the peak polarizations decrease monotonically with doping due to state mixing in the valence band. The latter is a general feature in semiconductors stating that emitting circularly polarized photons is not possible when $E_F$ is well below the split-off band \cite{OO_84}.

The results of Fig.~\ref{Fig:spectrum} elucidate the measured polarized electroluminescence (EL) from spin injected silicon \cite{Jonker_NaturePhysics07,Kioseoglou_APL09,Li_APL09}. The theory and experiment are consistent about the higher polarization of the TA related peak as well as for the polarization ratios between the TA and the TO dominated peaks. We find that the spin polarization of injected electrons in Refs.~\cite{Jonker_NaturePhysics07,Kioseoglou_APL09,Li_APL09} is about 27\% by comparing the measured and maximal achievable circular polarizations of the TA peak at 1105~meV. This relies on the ratio between the measured circular polarization of 3.5\% and the maximum attainable circular polarization of 13\% at 10$^{19}$~cm$^3$ p-type silicon (Fig.~\ref{Fig:spectrum}(e)) \cite{footnote1}. Another point of consideration in Refs.~\cite{Jonker_NaturePhysics07,Kioseoglou_APL09,Li_APL09} is the $\lesssim$10~ns Auger lifetime of electrons in the p-type substrate \cite{Dziewior_APL77}. The EL measurements probe those electrons that recombine radiatively within the effective lifetime. Quantitatively, about $\tau_A$/$\tau_r$ of the electrons that reach the substrate experience radiative recombination where $\tau_A$ and $\tau_r$ are, respectively, the effective (Auger dominated) and radiative lifetimes. The relatively short $\tau_A$ explains the mitigated spin relaxation effect in these measurements.

In conclusion, we have studied the luminescence in spin polarized silicon. Useful circular polarization degrees were provided for each of the phonon-assisted transitions. The antipodal behavior of optical phonons was shown to be responsible for the differences in the circular polarization of the luminescence peaks. The doping was shown to affect both the shape and polarization of the spectrum. Knowing the theoretical circular polarization values of complete spin polarization and comparing them with measured values are imperative in determining the spin polarization in silicon. These values are also instrumental in extracting the spin relaxation time or the spin injection efficiency across ferromagnet/silicon interfaces. Finally, the theory sets a basis for future studies of the circular polarization in indirect gap semiconductors due to strain (splitting the valleys and hole states), no-phonon and multi-phonon-assisted optical transitions.

We are indebt to Dr. G. Kioseoglou and Dr. B. Jonker for many fruitful discussions. This work is supported by AFOSR Contract No. FA9550-09-1-0493 and by NSF Contract No. ECCS-0824075.


\begin{thebibliography}{99}
\bibitem{OO_84}  Optical Orientation, edited by F. Meier and B. P. Zakharchenya (North-Holland, New York, 1984).
\bibitem{Zutic_RMP04}  I. \v{Z}uti\'{c}, J. Fabian, and S. Das Sarma, Rev. Mod. Phys. \textbf{76}, 323 (2004).
%
\bibitem{Lampel_PRL68} G. Lampel, Phys. Rev. Lett. \textbf{20}, 491 (1968). This poineer optical-orientation experiment used silicon.
\bibitem{Asnin_JETP76} V. M. Asnin, G. L. Bir, Y. N. Lomasov, G. E. Pikus, and A. A Rogachev, Sov. Phys. JETP, \textbf{44} 838 (1976)
\bibitem{Pikus_SPSS77} G. E. Pikus, Sov. Phys. Solid State, \textbf{19} 965 (1977).
\bibitem{Nastos_PRB07} F. Nastos, J. Rioux, M. Strimas-Mackey, B. S. Mendoza, and J. E. Sipe, Phys. Rev. B \textbf{76}, 205113 (2007).
%
\bibitem{Appelbaum_Nature07} I. Appelbaum, B. Q. Huang, and D. J. Monsma, Nature \textbf{447}, 295 (2007).

\bibitem{Jonker_NaturePhysics07}  B. T. Jonker, G. Kioseoglou, A. T. Hanbicki, C. H. Li, and P. E. Thompson, Nature Phys. \textbf{3}, 542 (2007).

\bibitem{Kioseoglou_APL09} G. Kioseoglou \textit{et al.}, Appl. Phys. Lett \textbf{94}, 122106 (2009). % A. T. Hanbicki, R. Goswami, O. M. J. van 't Erve, C. H. Li, G. Spanos, P. E. Thompson, and B. T. Jonker

\bibitem{Li_APL09} C. H. Li, G. Kioseoglou, O. M. J. van 't Erve, P. E. Thompson, and B. T. Jonker, Appl. Phys. Lett \textbf{95}, 172102 (2009).

\bibitem{Grenet_APL09}  L. Grenet \textit{et al.}, Appl. Phys. Lett \textbf{94}, 032502 (2009). %Spin injection in silicon at zero magnetic field; M. Jamet, P. Noé, V. Calvo, J.-M. Hartmann, L. E. Nistor, B. Rodmacq, S. Auffret, P. Warin, and Y. Samson

\bibitem{Zutic_PRL06} I. \v{Z}uti\'{c}, J. Fabian and S. C. Erwin, Phys. Rev. Lett. \textbf{97}, 026602 (2006). % Spin injection into silicon

\bibitem{Huang_APL08} B. Q. Huang, H. J. Jang, and I. Appelbaum,  Appl.
Phys. Lett \textbf{93}, 162508 (2008).

\bibitem{Dash_Nature09}  S. P. Dash, S. Sharma, R. S. Patel, M. P. de Jong, and R. Jansen, Nature \textbf{462}, 491 (2009).
%
\bibitem{Jansen_NatureMater10}  R. Jansen, B.-C. Min, and S. P. Dash, Nature Mater. \textbf{9}, 133 (2010).
491 (2009).
%
\bibitem{auxiliary} Auxiliary material.
%
\bibitem{Glembocki_PRL82} O. J. Glembocki and F. H. Pollak, Phys. Rev. Lett. \textbf{48}, 413 (1982); Glembocki, Ph.D Thesis, (1982).
%
\bibitem{Bednarek_PRL82} S. Bednarek and U. R\"{o}ssler, Phys. Rev. Lett. \textbf{48}, 1296 (1982).
%
\bibitem{Lax_PR61} M. Lax and J. J. Hopfield Phys. Rev. \textbf{124}, 115 (1961).
%
\bibitem{Cheli_PRB76} J. R. Chelikowsky and M. L. Cohen, Phys. Rev. B \textbf{14}, 556 (1976).
%
\bibitem{Weber_PRB77} W. Weber, Phys. Rev. B \textbf{15}, 4789, (1977).
%
\bibitem{Allen_PRB81} P. B. Allen and M. Cardona, Phys. Rev. B \textbf{23}, 1495 (1981).
%
\bibitem{Yafet_1963} Y. Yafet, in Solid State Physics, edited by F. Seitz and D. Turnbull (Academic, New York, 1963), Vol. 14, p. 76.
%
\bibitem{Cheng_PRL10} J. L. Cheng, M. W. Wu, and J. Fabian, Phys. Rev. Lett. \textbf{104}, 016601 (2010).
%
\bibitem{Wagner_PRB84} J. Wagner, Phys. Rev. B \textbf{29}, 2002 (1984).
%
\bibitem{footnote1}
By comparing the EL measurments in Refs.~\cite{Jonker_NaturePhysics07,Kioseoglou_APL09,Li_APL09} with the systematic photoluminescence measurements in Ref. \cite{Wagner_PRB84}, one finds that the correlated 1065~meV and 1105~meV peaks in Refs.~\cite{Jonker_NaturePhysics07,Kioseoglou_APL09,Li_APL09} originate from the $\gtrsim$10$^{19}$~cm$^3$ p-type wide substrate. The other measured correlated peaks at 1050~meV and 1090~meV relate to EL from the interface region rather than EL from the 300~nm overgrown $p$-$i$-$n$ diode region. The reason is that the doping levels in the overgrown region are smaller than in the substrate and as such the substrate band-gap is smaller. As a result EL from the overgrown diode region would involve photons of energy higher than 1065~meV.
%
\bibitem{Dziewior_APL77} J. Dziewior and W. Schmid, Appl. Phys. Lett. \textbf{31}, 346 (1977).

\end{thebibliography}
\end{document}